\documentclass[flushrt]{aastex} 

\shorttitle{Astrophysical Jets from Keplerian Disks}
\shortauthors{Vitorino, B.F., Jatenco-Pereira, V. & Opher, R.}

\def\bfit#1{\mbox{\boldmath $#1$}}
\begin{document}

\title{Numerical Simulations of Astrophysical Jets from Keplerian Disks
with Periodic Ejection.}

\author{B. F. Vitorino, V. Jatenco-Pereira and R. Opher}
\affil{Instituto de Astronomia, Geof\'{\i}sica e Ci\^{e}ncias Atmosf\'{e}ricas \\
Universidade de S\~{a}o Paulo \\
Rua do Mat\~{a}o 1226, 05508-090 S\~{a}o Paulo, SP, Brazil}

\email{braulo@astro.iag.usp.br, jatenco@astro.iag.usp.br, opher@astro.iag.usp.br}

\begin{abstract}
We present 2.5-dimensional time-dependent simulations of nonrelativistic and nonradiative
outflows from sinusoidally perturbed Keplerian accretion disks. A sinusoidal perturbation
is introduced in the velocity of the gas ejected from the surface of the disk into
a cold corona. In the simulations, the disk is a fixed boundary from which the gas is
ejected with a pulsed velocity. The maximum value of this velocity is taken to be a thousandth
of the local Keplerian disk velocity. It was found that for large periods, the structures in the
jet tend to fragment into smaller substructures. For small values of the period, the structures
tend to dissipate, while for medium values of the period, they tend to persist.

\end{abstract}
\section{Introduction}
In general, the production of astrophysical jets has been primarily explained by two possible
mechanisms: magnetic forces associated with an accretion disk \citep{konigl..99} and the
interaction between an accretion disk and a magnetized star, in the case of protostellar jets
\citep{shu99}.
The first mechanism is a disk-wind driven model, where the gas
is centrifugally flung out from the surface of a Keplerian disk when field lines thread the
disk at an angle of $60^\circ$ or less with respect to the surface.
When the gas reaches the local Alfv\'en velocity, the ($\bfit{J_p} \times \bfit {B_\phi}$) term
becomes dominant, providing the collimation for the outflow. 
The second mechanism is the X-wind model, where the wind emanates near the region where
the disk corotates with the central magnetic star. It is assumed that some magnetospheric lines,
connecting the central star and the disk, are partially open. Excess angular momentum
brought in by the disk is extracted by the magnetocentrifugally driven wind. The wind emanates
from the inner edge of the disk and flows out along the open magnetic field lines.
Since the first mechanism seeks to explain astrophysical jets independently of the detailed
nature of the central object, from active galactic nuclei (AGNs) to young stellar objects
(YSOs), it has the possibility of being a universal model.

Recent theories and models of astrophysical jets have mainly tried to provide
a natural explanation for common features. These include the episodic outbursts and knots
that are observed.

Numerical simulations of axisymmetric magnetized flows from accrection disks have previously
been made using different models. \cite{shib..86} and \cite{stone..94} followed the internal
dynamics of the disk. \cite{bell..95} and \cite{usty..95} treat the disk as a boundary
condition at the base of the wind, without following its internal dynamics, and assume it to
be in Keplerian equilibrium. Following the latter model, Ouyed \& Pudritz (1997a, b; 1999)
made 2.5-dimensional, high-resolution, numerical magnetohydrodynamic (MHD) simulations of the
onset and collimation of outflows from the surface of a Keplerian accretion disk around a
central object. Using two different topologies for the magnetic field that threads the disk and
the corona, they obtained steady or episodic jets, depending on the values of the constant
(in time) ejection velocity from the disk to the corona. For the smallest values of the
ejection velocity, their results showed episodic production of knots near the surface of
the disk which propagate along the jet axis. They argued that these knots arose due to MHD
shocks. 

\cite{krasno..99} made time-dependent, axisymmetric simulations of the
magnetocentrifugal model jet formation. Their system consisted
of a Keplerian disk and a central object. The main subtlety resides in the treatment of the
disk-wind boundary, the disk surface. Some, but not all, of the flow variables can be fixed
on this boundary. The boundary conditions imposed on the disk surface were limited to those
necessary and sufficient to take into account information propagating upstream from the fast
and Alfv\'en critical surfaces, avoiding overdetermination of the flow and the production of
impulsive accelerations.
They obtained a cold and steady jet, launched smoothly from the Keplerian disk and collimated by
the magnetocentrifugal mechanism. The collimation is observed both in the shape of the field
lines and in the density profiles, which become cylindrical, and thus jetlike, at large
distances
along the rotation axis. The structure of the flow is insensitive to the density or flow
speed at the injection surface, as long as the mass flux is kept constant. As explained by
the authors, steady-state jets obtained in earlier
studies, using an essentially incomplete treatment of the disk boundary conditions, are
qualitatively unchanged when this deficiency is rectified. It remains to be shown, however,
whether the same is true for nonsteady jets.

\cite{frank..2000} made a series of simulations intended to address the issue of
MHD jet propagation. Whereas some studies have used ad hoc initial conditions, they injected
flows into a computational grid derived from models of collimated jets driven by
magnetocentrifugal launching. The initial configurations in the jet were taken directly from
the solution of the force balance perpendicular (the Grad-Shafranov equation) and parallel
(the Bernoulli equation) to the magnetic surfaces generated by a magnetized rotator. Their
simulations followed the evolution of jets composed of helical fields, embedded in hypersonic
plasmas whose density and velocity varied with radius. They studied the propagation of jets
driven by: (1) a purely Keplerian rotator; and (2) a Keplerian rotator with a sub-Keplerian
boundary layer. Both rotators are exterior to a solid body rotator. Studying the adiabatic case
for the last model (their multicomponent model) they obtained a jet in which strong
instabilities developed in the core of the jet. Once the core is exposed, periodic pinches
can be seen in
the beam. Their stability analysis indicates that the origin of these periodic pinches could
be current driven instabilities. The pinching instabilities on the axis have a wavelength of
approximately half of the jet radius. However, the beam also has a second characteristic
wavelength which runs along the surface of the core. This feature, which appears as an envelope
encompassing the shorter wavelength modes, has a wavelength of approximately three jet radii.
As explained by the authors, these results suggest that these instabilities could be at the
origin of the knotty structure of a large number of jets as seen, for example, in HL Tau, HH1,
HH30, and HH34.       

Time-dependent MHD calculations of a modified version of the X-wind model, considering the
star-disk field interaction, shows twisting of closed stellar field lines, current sheet
formation, reconnection, and ejection of magnetic island with the plasma heated to
$\leq 10^8$ K, consistent with X-ray flare observations \citep{hayashi..96,goodson..97}. 

All the above previous studies assumed constant velocity injection from the disk. Here, we
extend the previous investigations and study the effect of non-steady injection, in
particular, periodic injection.

Adopting the disk as a boundary condition, we investigated how periodic perturbations at the
disk surface can alter the formation and the propagation of the magneto-centrifugal jet.
In the FU-Orionis phase of a protostar, for example, we know that there is a periodic ejection
of material observed in the X-ray range. We do not include the physics of the disk that could
generate a pulsed gas ejection. We assume that there is a periodic mass loading at the base of
the corona of the disks, which introduces a periodic ejection of gas from the disk surface.
Whereas Ouyed \& Pudritz (1997a, b; 1999) used a constant ejection velocity from the disk, we
study the temporal dependence of this velocity.
We made 2.5-dimensional numerical simulations using
the ZEUS-3D code \citep {stone..92a,stone..92b} in the axisymmetry option. A sinusoidal
perturbation was introduced in the ejection velocity as a temporal function. Various values
for the periods in the range from 10 to 80 rotation periods of the innermost radius of the
accretion disk ($r_i$), were used in the simulations. The maximum value for the perturbation
was chosen. This value is roughly equal to that which would produce a steady state jet if the
gas were continuously ejected from the disk \citep*[hereafter OP97a]{ouyed..97a}. The
luminosity of the protostellar T Tauri stars (which have accrection disks) fluctuates rapidly
with large amplitude. Similarly, quasars, which are also assumed to have accretion disks,
also have luminosities which fluctuate rapidly with large amplitude. Therefore, we introduced
in our simulations velocity perturbations that are a large percentage of the injection
velocity. A region near the central object, $80 r_i$ in the $z$-direction and $20 r_i$ in
the $r$-direction, was simulated.

Using different values for the period of the sinusoidal perturbation, we obtained the formation 
of regularly spaced structures along the jet axis. For small values of the period, the
structures tend to dissipate along the jet axis; for medium values, they
tend to persist and for large values, tend to fragment into smaller
substructures. 

In \S 2 the basic equations that are solved are presented.
The model is described in \S 3 and the numerical results are presented and
discussed in \S 4. Finally, in \S 5 we present a summary and the final conclusions.

\section{Basic equations \label{2}}

Adopting cylindrical coordinates ($r$, $\phi$, $z$),
we placed the central object at the origin and took the $z$-axis to be perpendicular
to the disk. The surface of the disk thus lies in the $z=0$ plane of our
coordinate system.

The equations that describe the conservation of mass, momentum, energy and
magnetic flux are: 
\begin{equation} \label{masscon}
\frac{\partial \rho }{\partial t}+{\bf \nabla} \cdot \left( \rho \bfit{v}\right) =0,
\end{equation}

\begin{equation} \label{momcon}
\rho \left[ \frac{\partial \bfit{v}}{\partial t}+\left( \bfit{v}\cdot {\bf \nabla} \right)
\bfit{v}\right] +{\bf \nabla} \left( p+p_A\right) +\rho {\bf \nabla} \Phi -\bfit{j}\times
{\bfit B} =0,
\end{equation}

\begin{equation} \label{enegcon}
\rho \left[ \frac{\partial e}{\partial t}+\left( \bfit{v}\cdot {\bf \nabla} \right)
e\right] +p\left( {\bf \nabla} \cdot \bfit{v}\right) =0,
\end{equation}

\begin{equation} \label{indeq}
\frac{\partial \bfit{B}}{\partial t}-{\bf \nabla} \times \left( \bfit{v}\times \bfit{B}
\right) =0,
\end{equation}

\begin{equation} \label{divb}
{\bf \nabla} \cdot \bfit{B}=0,  
\end{equation}
with $\rho $ the gas density, $\bfit{v}$ the velocity, $\bfit{B}$ the magnetic field,
$p$ the gas pressure, $e$ the internal energy, $\bfit{j}={\bf \nabla} \times \bfit{B}/4\pi$
the electric current density, and $\Phi =-GM/\left( r^2+z^2\right) ^{1/2}$ the gravitational
potential.

A polytropic gas $p=K\rho ^{\gamma}$ with the polytropic index $\gamma=5/3$ is assumed. Thus,
ignoring radiative transfer effects, we do not solve the energy equation (\ref{enegcon}).
We also assume that $p_A$ is an Alfv\'enic turbulent pressure. This provides, together with
the thermal pressure, the main support against gravity for the cold corona. Here $p_A=p/\beta
_T$, where $\beta _T$ is the ratio of the gas pressure to the Alfv\'enic turbulent pressure,
assumed to be constant (\citeauthor{ouyed..97a}).

All parameters are given in units of their values at the innermost
radius of the accretion disk $r_i$.
Thus, the normalized equation of motion (eq. [\ref{momcon}]) is 
\begin{equation} \label{momconnor}
\left[ \frac{\partial \bfit{v}^{\prime }}{\partial t}+\left( \bfit{v}^{\prime }\cdot
{\bf \nabla} ^{\prime }\right) \bfit{v}^{\prime }\right] =-\frac 1{\delta _i}\frac{{\bf \nabla}
^{\prime }\left( p^{\prime }+p_A^{\prime }\right) }{\rho ^{\prime }}-{\bf \nabla}
^{\prime }\Phi ^{\prime }+\frac 2{\delta _i \beta_i}\frac{\bfit{J}^{\prime }\times
\bfit{B}^{\prime }}{\rho ^{\prime }}, 
\end{equation}
where the dimensionless variables are: $r^{\prime }=r/r_i$, $z^{\prime }=z/r_i$, $v^{\prime }
=v/v_{K,i}$, $\rho ^{\prime }=\rho /\rho _i$, $p^{\prime }=p/p_i$, $B^{\prime }=B/B_i$,
$\Phi ^{\prime }=-1/\left( r^{\prime 2}+z^{^{\prime }2}\right) ^{1/2}$ and
$\nabla ^{\prime }=r_i\nabla $. The parameter $\delta _i=\rho_iv_{K,i}^2/p_i$ is the
ratio of the Keplerian to the thermal energy density and $\beta _i=8\pi p_i/B_i^2$ is the ratio
of the gas to the magnetic pressure in the corona at $r_i$. At $r_i$, the Keplerian speed 
is $v_{K,i}=\left( GM/r_i\right) ^{1/2}$ and $B_i$ is the poloidal field at $r_i$.
For convenience, we drop the primes and refer only to the normalized variables.
Time is given in units of $t_i=r_i/v_{K,i}$, so that the dimensionless time
is
\begin{equation} \label{time}
\tau =\frac t{t_i} .
\end{equation}

\section{The Model\label{3}}

In this model, the disk is a fixed boundary condition in pressure
equilibrium with the corona. At the center there is a point mass
that represents a star or a black hole. An initial potential (zero current) poloidal
field threads the disk and the corona.

The computational
domain defines the spatial region where the MHD equations are time evolved.
ZEUS-3D, a grid code, divides the computational domain into cells, 
called ``active" zones. Finite differencing the evolution near the
boundaries of the grid requires that values for the dependent variables be specified
beyond the computational domain. Therefore, at each boundary, ``ghost" zones are added.
Values for the dependent variables in the ghost zones are specified, using boundary
conditions appropriate to the geometry and the physics of the problem being solved.
Thus, the evolution equations are not solved for the ghost zones. 

We use inflow boundary conditions at the disk surface and outflow boundary conditions on the
remaining boundaries. Reflecting boundaries are used along the axis of symmetry, which coincides
with the disk axis. The normal component (radial component) of the velocity and the magnetic field are set to zero on the boundary and reflected into the ``ghost" zones. In addition, the
3-component (toroidal component) of the velocity and magnetic field is also inverted along the
axis of symmetry into the ``ghost" zones. The initial setup and boundaries are shown in Figure \ref{f1} (Fig. 2 of \citeauthor{ouyed..97a}).

\placefigure{f1}

In the corona, the initial toroidal field is zero. Magnetostatic
equilibrium is numerically maintained to assure that an outflow can only arise
due to the magnetocentrifugal mechanism.
Resolving equation (\ref{momconnor}) for magnetostatic equilibrium, we
obtain the expression for the coronal density distribution,
\begin{equation} \label{cordens}
\rho =\left( 1/\sqrt{r^2+z^2}\right) ^{3/2} ,  
\end{equation}
which is used in our numerical simulations.

Since the surface of the accretion disk is a fixed boundary condition, there is no back
reaction by the jet. We are, thus, time evaluating the coronal region ($z\geq 0$) only where
the jet arises. However, we fix the density at the base of the corona (first column of cells),
assuming that the disk is a suitable dynamic source of matter. The density distribution of
the disk surface has the form \begin{equation} \label{diskdens}
\rho _d=\eta _ir^{-3/2} ,
\end{equation}
where $\eta _i=\rho _d/\rho _0$ is the ratio between the disk density ($\rho _d$) and
the density at the base of the corona ($\rho_0 \equiv \rho(z=0)$).

The corona and the disk are threaded by the poloidal magnetic field in a potential
configuration ($\bfit{J}=0$), described by the stream function,
\begin{equation} \label{streamfunc}
\Phi_m = \sqrt {r^2+\left( z_d+z \right)^2}- \left( z_d+z \right),
\end{equation}
where $z_d$ is the dimensionless disk thickness.
The field components are given as
\begin{equation} \label{fieldcomp}
B_z= \frac {1}{r}\frac {\partial {\Phi_m}} {\partial {r}} \;\;\; {\mathrm{and}} \;\;\; 
B_r= - \frac {1}{r}\frac {\partial {\Phi_m}} {\partial {z}} .  
\end{equation}

Due to the torsion of the poloidal field at the disk surface, a toroidal magnetic field
is produced at the base of the corona, which could propagate into the
disk. We thus expect that the disk has a nonzero toroidal field,
\begin{equation} \label{btor}
B_\phi =\frac{\mu _i}r ,
\end{equation}
where the input parameter $\mu _i=B_{\phi i}/B_i$ is the ratio of the toroidal to poloidal
magnetic field at $r_i$ .
There are three important initial timescales defined at the
disk's surface: the Kepler time $\tau _K$, the Alfv\'en time $\tau _A$ 
and the magnetic braking time $\tau _B$. Thus, we have (e.g., eq. [4.64] of
\citeauthor{ouyed..97a})
\begin{equation}
\left( \tau _K,\;\tau _A,\;\tau _B\right) =\left( 1, \;\; \sqrt{\delta _i\beta_i/2}, \;\; \eta
_i\sqrt{\delta _i\beta _i/2}\right) .
\end{equation}

To estimate the turbulent pressure needed to support the corona for a given
gas pressure, we use the equation for the ratio of the turbulent to magnetic pressure $\beta _T$
\begin{equation} \label{betaturb} 
\beta _T= \gamma / \left( \left( \gamma -1\right) \delta _i-\gamma \right),
\end{equation}
where, as in equation (\ref{momconnor}), $\delta_i$ is the ratio of the Kepler to thermal
energy density.
The gas is injected from the disk into the corona at a velocity $\bfit{v}_p=f_v
v_K \bfit{B}_p/B_p$, where $v_K= \left( 1/r \right)^{1/2}$ is the local Keplerian velocity at
the disk surface (for $r \geq 1$)\footnote{For the radial region where the disk does not exist ($r < 1$ 
(i.e., $r < r_i$)) we set
$\bfit{v}_p$ to zero.}, $\bfit{B}_p$ is the local
poloidal magnetic field that threads the disk, and $f_v$ is the temporal function that
describes the temporal dependence of the injection velocity. In our simulation we introduced
a pulsed injection velocity with a sinusoidal form,
\begin{equation} \label{vinj}
f_v = v_{inj0}\left[1+\sin \left( 2\pi \tau /T\right)
\right] ,
\end{equation}
where $T$ is the period of the perturbation, $v_{inj0}$ is an input parameter and $\tau$ is
given by equation (\ref{time}).

We, thus, have six free parameters in our simulation: 
$\delta _i$, $\beta _i$, $\eta _i$, $\mu _i$, $v_{inj0}$ and $T$. 

\section{Numerical results and discussion \label{4}}

We used the following parameters to investigate the time evolution of a jet from a Keplerian
accretion disk with pulsed gas injection (similarly to \citeauthor{ouyed..97a}):
\begin{equation}
\left( \delta _i, \beta _i, \eta _i, \mu_i, v_{inj0}\right) = \left(
100.0,1.0,100.0,1.0,10^{-3}\right) ,
\end{equation}
implying the initial timescales
\begin{equation}
\left( \tau _K,\tau _A,\tau _B\right) = \left(1.0,7.07,707.1\right) .
\end{equation}
Five simulations were made. In simulation A, there was no perturbation; in simulations
B to E, we introduced perturbations with various values for the period $T$.
The simulations were made in the domain $\left( z,r\right)=\left(80,20\right)$
with a resolution of $\left( 120,45\right)$ for 400 time
units $\left( <\tau _B\right)$. 
Along the z-axis, 30 zones span a distance of $10r_i$ outwards from the disk
surface. An additional 90 ratioed zones span an additional distance of 
$70r_i$, with each subsequent zone increasing in size by a factor of 1.017.
Twenty uniform zones span an initial five disk
radius in the radial direction. Outside of these uniform grid zones, 25 zones are
ratioed, with each subsequent zone increasing in size by a factor 1.065.

Figure \ref{f2} shows the initial conditions in the atmosphere. The density
distribution (eq. [\ref{cordens}]) is shown at the top and the initial magnetic
(potential) configuration, at the bottom of the figure. Initially, the velocity and the toroidal
magnetic field are zero.

\placefigure{f2}

The results of the simulations are shown in a series of figures that are described below.
Figures \ref{f3} to \ref{f7} show the distributions of density, poloidal
velocity (vectors), poloidal and toroidal magnetic fields and toroidal velocity at 400 time
units for the simulations A to E, respectively. The contour lines are isocontours in the
$r-z$ plane.

In simulation A (Fig. \ref{f3}), no perturbation was introduced in the gas injection.
Results in good agreement with those of \citeauthor{ouyed..97a} were obtained. From the density
distribution, shown at the top of this figure, we observe the formation of a highly collimated
steady state jet along the z-axis. The collimation is due to the strong radial pinch force of
a dominant toroidal field (fourth panel from top to bottom) that propagates into the corona.
We obtain a maximum value for the poloidal velocity (second panel from top to bottom) on the
order of the Keplerian velocity at $r_i$.

\placefigure{f3}

A sinusoidal perturbation in the gas injection velocity (eq. [\ref{vinj}]) at the disk surface
was introduced in the other four simulations. As we are interested in studying the
behaviour of the jet subject to this perturbation at its base, we chose a progressive range
for the period T. In simulations B to E, we use $T=10,20,40,80$, respectively.

Along the jet axis in simulation B for $T=10$ (Fig. \ref{f4}), we observe the formation of
regularly spaced structures up to about $z=50r_i$. The distance of separation between two
consecutive structures is about $5r_i$ (i.e., $\sim \left ( T/2 \right ) r_i$). The
axial extent of the regularly spaced structures advance with the flow speed until it reaches
the limiting distance of about $50r_i$. This pattern persists even for times greater than
400 time units.

\placefigure{f4}

Our results were obtained using ratioed zones that allow for a smaller number of calculated
zones and, therefore, a poorer resolution as compared with \citeauthor{ouyed..97a}. To verify
the sensitivity of our results to this poorer resolution, we made simulation C using only
uniform zones at a resolution close to that used in the uniform zone region in the poorer
resolution case. This simulation was run in the domain $(z,r)=(80,20)$ with the resolution of
$(250,80)$ uniform zones. For the others cases (D -- E), we used the poorer resolution.

Regularly spaced structures along the whole jet (Fig. \ref{f5}) are seen in simulation C,
where $T=20$. These structures persist during the entire simulation, even for times greater
than 400 units. The distance of separation between two consecutive structures is about
$11r_i$, or roughly $\left ( T/2 \right ) r_i$. We obtained these same results when
using ratioed zones in the poorer resolution case.

\placefigure{f5}

We observe the formation of  regularly spaced structures
along the entire jet as well as their fragmentation in simulation D for $T=40$.
Few minor structures are observed between the major structures (Fig. \ref{f6}: top).
Fragmentation of the structures can be seen in the isocontours of the toroidal
velocity in the bottom panel of Figure \ref{f6}. These structures persist during
the entire simulation, even for times greater than 400 units. The distance
of separation between two major structures is about $22r_i$, i.e., roughly
$\left ( T/2 \right ) r_i$.

\placefigure{f6}

In simulation E for $T=80$, we observe the formation of 
regularly spaced structures along the entire jet (Fig. \ref{f7}: top), with more pronounced
fragmentation clearly seen in the plot of the toroidal velocity (Fig. \ref{f7}:
bottom). These structures persist during the entire simulation,
even for times greater than 400 units. The distance of separation between two
major structures is about $40r_i$, i.e., $\sim  \left ( T/2 \right ) r_i$.

\placefigure{f7}

The axial distributions, for all the simulations, of density, axial velocity, toroidal
velocity and toroidal magnetic field along a cut at $r=5r_i$, parallel to the disk's axis
at $400$ time units, are shown in Figure (Fig. \ref{f8}).
These quantities are normalized to their maximum values along this cut. We note from the
axial velocity distribution (dotted line),  that the flow is rapidly accelerated along the
field line from the disk to roughly $z=15r_i$. After this point there is a ``Hubble flow"
constant acceleration, with $v_z$ proportional to $z$ for simulation A. However, for
simulations B to E this constant acceleration character is modified, reflecting the pulsed
nature of the jet. From the toroidal magnetic field distribution (dashed-dotted line) we
observe the effect of torsion of the poloidal field lines due to the rotation of the disk.
The regular structures, in simulations B -- E, have their minimum mass concentration
(solid line) at the maximum values of toroidal magnetic field (dashed-dotted line), reflecting
jet constriction by this field. The shift between the toroidal magnetic field and the
density in the pinched structures is a general result of long standing (e.g., \cite{chan..80})
and it has been associated with the periodic oscillations caused by the pinch instability due
to the toroidal magnetic field.
 
\placefigure{f8}

For simulations A to E at
$\tau =400$, the axial and radial mass fluxes are shown in Figure \ref{f9}. 
The mass flux is defined as

\begin{equation} \label{massflux}
\dot{m}= \int \rho \, \bfit{v} \cdot \bfit{dA} ,
\end{equation}
where $A$ is the area.
The axial mass flux\footnote{The computation of the axial mass flux involves computing the product 
of $v_z$ and the area
of each annulus for each radial zone $[\pi (r^2_{j+1} - r^2_j)]$, where $j$ is an active zone
and $r_j$ refers to the position of the face or edge of the zone. At each $z$, the axial mass flux
is computed from the sum of the axial mass flux contributions of all 45 radial zones.} refers
to the quantity of mass per unit time which passes through a circular area $A=\pi r^2$ and is
calculated for each $z$, using $r=20r_i$; the radial mass flux\footnote{The radial mass flux at 
each $z$ involves computing the product of $v_r$ and the area of the
annulus for a given radial zone $[2 \pi r_j (z_{k+1} - z_{k})]$, where $j$ and $k$ define each
active zone, and $r_j$ and $z_k$ refers to the face/edge position of the zone, along the $r$
and $z$ direction, respectively. In particular, we computed the radial flux at each $z$ using
$r=20r_i$.} 
refers to the quantity of mass per unit time which passes through the area
$A=2 \pi r z$ and is shown for each z, using $r=20r_i$. These fluxes, given in units of
$\dot{m}_i=2 \pi r_i^2 \rho_i v_{Ki}$, are important for the detection of the
decollimation of the jet. Their values agree with the results of \citeauthor{ouyed..97a}. 
At the time $\tau =400$, the axial fluxes at $z=80r_i$ are comparable to those
at $z\sim 0$; the small difference owing to decollimation of the jet.
As the axial mass flux decreases, the radial flux can be seen to increase by
roughly the same amount.
We can see the pulsed nature of the jet (simulations B to E) from the axial mass fluxes (solid
lines) which oscillate around their average values. The radial mass fluxes in simulations B to E
are similar to those of simulation A (without perturbation).
In simulation B, the axial mass flux curve is not very modified by the density structures, as
compared with the case for which there is no perturbation (simulation A);
in simulations C to D, the axial mass flux curves are strongly modified by the
density structures. Their average values, however, do not change.

\placefigure{f9}

Figure \ref{f10} shows the sonic, Alfv\'enic and magnetosonic Mach numbers for all the
simulations at $\tau =400$,
in a cut at $r=5r_i$ along the $z$-axis. In the computation of the sonic Mach numbers, we use
only the thermal pressure $p$ in the sound speed. To compute the magnetosonic Mach number, we
use the effective sound speed which includes the thermal and the Alfv\'enic turbulent
pressure, $p_A$\footnote{The magnetosonic speed is defined as $\left ( c_s^2 + v_{Az}^2 \right )
^{1/2}$ where $c_s = \left ( \gamma \left ( p + p_A \right ) / \rho \right )^{1/2}$ is the
effective sound speed and $v_{Az} = B / \left ( 4 \pi \rho \right )^{1/2}$ is the axial
Alfv\'en velocity.}. We note in Figure \ref{f10} that the jet is already supersonic and
super-Alfv\'enic near the base of the corona and super-magnetosonic at $z\sim 4r_i$. Only
where density structures exist are the Mach numbers (Fig. \ref{f10}) modified. Their
median values, however, remain roughly the same. The maximum values of Alfv\'enic Mach numbers
(dashed-dotted line) are due to the decreased axial magnetic fields. Negative values correspond
to a reversal of the axial magnetic field.

\placefigure{f10}

\section{Summary and Conclusions \label{5}}

We made numerical 2.5-dimensional simulations of astrophysical jets produced from
Keplerian disks subject to a sinusoidal perturbation of the gas
injection velocity from the disk into the corona. Ratioed zones were used in order
to obtain better resolution near the region of the generation of the jet with a poorer
resolution outside.

Five simulations were made with different values of the period of the
sinusoidal perturbation. No perturbation was introduced in
simulation A.
In simulations B to E, we used periods $T=10, 20, 40$ and $80$, respectively, in units of
$t_i=r_i/v_{Ki}$, where $r_i$ is the innermost radius of the disk and $v_{Ki}$, its Keplerian
velocity.

For simulation A, where there was no perturbation, we obtained a steady state  highly collimated
jet with a maximum axial velocity of the order of $v_{Ki}$ after a time
$\tau =400$. In a cut at $r=5r_i$ parallel to the disk's
axis, the jet is supersonic and already super-Alfv\'enic  near the base of the corona and
super-magnetosonic at $z\sim 4r_i$. These results agree with those of \citeauthor{ouyed..97a},
who used better resolution.

Introducing a sinusoidal perturbation in the gas injection velocity, the
results common, to simulations B to E, are:

1. The formation of regularly spaced structures
with a separation distance of roughly $\left ( T/2 \right ) r_i$ along the jet axis.

2. The toroidal magnetic field is out of phase ($\sim 90^\circ $) with the
density curve along the jet axis.

3. The axial mass fluxes are modified by the periodic density structures, although the average
remains roughly the same, as compared with the same curve for the
case of no perturbation (simulation A). The axial fluxes at $z=80r_i$ are comparable
to the axial fluxes at $z\sim 0$. 

4. The radial mass fluxes are similar to those of simulation A (where there was no
perturbation). However, differences arise when we examine a cut at $r=5r_i$. There is
oscillatory behaviour in some of the simulations at this cut.  

5. In a cut at $r=5r_i$ the Mach number distributions are modified only where density
structures exist. The median values, however, remain roughly the same, and are similar to
simulation A.

The results that are different in some of these simulations are:

a. In simulation B ($T=10$) the structures persist only up to an axial distance of
$z\simeq 50r_i$.

b. In simulations D ($T=40$) and E ($T=80$), fragmentation of the structures occur.

These simulations, introducing periodic perturbations in the accretion disk, produced
collimated jets with a wavelength of $\left ( T/2 \right ) r_i$, where $T$ is the temporal
period of the perturbation. The creation of structures with a specific distance separation
was determined by the period $T$ of the gas ejection from the disk surface. For example, we
obtained a jet with a distance separation of about $11r_i$ associated with a period of 20 time
units for the perturbation. However, for different values of $T$, an interesting effect occurs:
the structures tend to dissipate at large axial distances for smaller values of $T$ and
to fragment into smaller structures for greater values of $T$. 

Using this same numerical approach, Ouyed \& Pudritz(1999) obtained episodic jets, using
steady gas injection from the disk with velocities which were smaller than $10^{-3}
v_{K,i}$ which
we used. For injection velocities equal to this value, they obtained a highly collimated
steady state jet\footnote{They obtained steady state jets with velocities equal to or greater
than $5 \times 10^{-4}v_{K,i}$, as well as episodic jets with different distances of separation for
smaller velocities, using the same free parameters that we used.} with features similar to
ours, with no perturbation in the velocity field of the disk. When we introduced
perturbations with specific features in our steady state jet, our results showed the tendency
to produce regular structures with separation distances $\sim 11r_i$, similar to the case where
OP99 used a smaller velocity of $10^{-4}v_{K,i}$. 

\cite{gardiner..2000}, using a magnetocentrifugal approach for the jet launching, studied an
analytical one-dimensional model, in which they investigated the behaviour of a jet with a
variable
output speed. They inferred that a pulsing jet with an embedded helical magnetic field will
inevitably develop a periodic field structure, consisting of rarefied regions (where the helix
is ``combed out", resulting in a more poloidal geometry) alternating with denser, toroidally
dominated knots. To confirm these predictions, they performed axisymmetric (2.5-dimensional)
numerical simulations of a radiative MHD jet. They used conditions appropriate for protostellar
jets and introduced a pulsed velocity with a sinusoidal form at the jet nozzle. Their results
showed that pulsing jets with initially helical fields will evolve into an alternating
poloidal-toroidal-dominated field geometry. The strongest toroidal field are confined to
the dense knots, while predominantly poloidal fields are in the rarefied regions.

Both our results and those of \cite{gardiner..2000} show that episodic jets are produced
by introducing a pulsed mass load at the base of the jet. In particular, in our simulations, as
well as in the study made by \cite{ouyed..97b}, knot formation is associated with an
anticorrelation of density and toroidal field strength. However, it should be noted that
our study is concerned with the near-field region (i.e., the field region near the origin of
the jet), whereas other studies, such as \cite{gardiner..2000}, are concerned
with the far-field region.

There is currently some debate over the origin of the knots in protostellar flows. It remains
unclear as to whether the knots are due to the pulsing of the source of the jet or to
instabilities in the
beam. Our results show that for a given constant ejection velocity from the disk surface which
produces a steady state jet, an episodic jet is produced when the given ejection velocity is
pulsed. We showed that the temporal and spatial knotty structure of the jets depends on
the period of the velocity perturbation at the base of the jet.

Ouyed \& Pudritz (1997b, 1999) rule out the possibility of a
pinch MHD instability causing the knotty structure. This instability depends upon the strength
of $\beta$ and no such dependence was found by the authors. They found that only for $\beta_i$
around unity, were knots produced. \cite{birkinshaw..90} noted that the Kelvin-Helmholtz (KH) 
instability does not exist if the beam speed
is less than the Alfv\'en speed. \cite{hardee..99} showed that KH growth rates increase as the
magnetosonic Mach number decreases, provided the jet is super-Alfv\'enic and the magnetized jet is
nearly stabilized to the KH instability when the jet is sub-Alfv\'enic. \cite{hardee..2002} found
that the jets can be stalilized to the KH helical and higher order asymmetric normal modes provided
the velocity shear, $\Delta_u \equiv u_j - u_e$ between the jet and the external medium is less than a
``surface" Alfven speed, $v_{AS} \equiv \left[ \left( \rho_J + \rho_E \right) \left( B_J^2 +B_E^2 \right)
/ \left( 4 \pi \rho_J \rho_E \right) \right] ^{1/2}$, where
$\rho_J (\rho_E)$ is the jet (external medium) density and $B_J (B_E)$ is the jet (external medium)
magnetic field. \cite{ouyed..2003} showed that the jet beyound the Alfv\'en surface becomes
unstable to nonaxisymmetric KH instabilities, in three-dimensional simulations of jets from Keplerian
accretion disks. 

We also made a number of simulations using values of $\beta_i$ different from unity. 
Figure \ref{f11} shows the results of one of these simulations, a periodic perturbation at the disk 
surface, using a period of $T = 20$. We plotted the isocontours of the density. The results for 
$\beta_i = 0.1$ are shown at the top of the figure. For this case, we see a smoothing of the 
structures along the jet. The isocontours of the density oscillate slightly near the jet axis. At the 
botton of this figure, the results for $\beta_i = 10$ are shown. Here we
observe the formation of structures from the jet axis up to a distance of about $z = 35r_i$.
The region near the jet axis shows irregular structures and is less well defined (compare with Figure \ref{f5}, where $\beta_i = 1$).

Our results are consistent with those of Hardee \& Rosen (1999, 2002) and \cite{ouyed..2003}.
This is shown in the sequence of simulations in Figures 5 and 11. When $\beta_i$ is small (strong
magnetic fields, and the flow is sub-Alfv\'enic), there is a not much growth of the instability.
The super-Alfv\'enic case in which $\beta_i$ is close to unity (the flow speed is close to the 
Alfv\'en speed) is more unstable.

\placefigure{f11}

We simulated a region very near the central object ($20r_i \times 80r_i$). It is
interesting to evaluate the spatial dimensions involved in our simulations. For a protostar
of mass $0.5 M_{\odot}$, taking $r_i = 0.04$ AU for the inner radius of the disk, we have
$80r_i \sim 10^{-5}$ pc. For a black hole of $10^8 M_{\odot}$ (the
Schwartzschild radius is $\sim 1$ AU), assuming $r_i = 20.6$ AU, we have $80r_i \sim 10^{-2}$
pc.  A distance of separation of $\sim 11r_i$ of the structures
corresponds, for the protostar case, to $\sim 10^{-6}$ pc and for the black hole case, to
$\sim 10^{-3}$ pc. These distances are much smaller than those between the Herbig-Haro
objects observed in protostellar jets (about $10^{-2}$ pc). They are also much smaller
than the distances observed between the emission knots, frequently found in AGN jets. In
contrast to other simulations (e.g., Frank et al.(2000)), we obtain structures very near
the formation region of the jets, involving distances not yet resolved by current instruments.

We can not say, at the present moment, whether or not the origin of the knotty structures seen
in many jets at large distances from the source are due to the structures
observed in the present simulations for regions close to the source. Further studies are
needed to investigate this question.

The instabilities seen by Ouyed \& Pudritz (1997b, 1999), producing the knotty
structures, occur at the fast magnetosonic surface and propagate downstream.
These instabilities, as explained by the authors, are due to MHD shocks. Some
conditions in the knot generating region (in particular, for high $\bfit{B}_{\phi}/
\bfit{B}_z$) seem  to point towards a scenario in which the knot spacings could
possibly be identified with the resonant wavelength pinch mode. As noted above, 
Ouyed \& Pudritz (1997b, 1999) rule out the possibility of a
pinch MHD instability causing the knotty structures since it has a strong $\beta$ dependence,
which was not found by the authors. The authors only observed knotty structures when $\beta$
was around unity.

We believe that the knotty structures that we obtained in our simulations are due to the
association of the pulsing mass load at the base of the jet and the instability due to MHD
shocks, discussed in the studies of Ouyed \& Pudritz (1997b, 1999). Our results indicate that a
particular frequency of the pulsed velocity can excite the instability and produce persistent
knotty structures along the jet.

\acknowledgments 
The authors would like to thanks the editor, Steven Shore, and the anonymous referee for helpful
comments. B. F. V. acknowledges the financial support of the Brazilian agency FAPESP
under grant 95/99122. V. J. P. and R. O. would like to thank the Brazilian
agency CNPq for partial support. We would like to thank the LCCA-USP for the possibility
of using the CRAY J-90 computer in our simulations and especially the staff for their
invaluable help. R. Ouyed and especially C. Fendt contributed to this paper with encouraging and 
valuable discussions, for which we are grateful. The authors also would like to thank the project
PRONEX (No. 41.96.0908.00) for partial support. R. O. would like to thank the Brazilian
agency FAPESP for partial support under grant 00/06770-2. V. J. P. would like to thank
FAPESP for partial support under grant 2000/06769-4.

\clearpage 
\figcaption[f1.ps]{Setup of the numerical simulations. Ghost zones ($z<0$)
define the surface of the Keplerian disk, while active zones (which are
evolved in time) define the corona. The poloidal magnetic field continues
smoothly into the disk surface (Fig. 2 of OP97a).
\label{f1}}
\figcaption[f2.ps]{Initial conditions in the atmosphere: density
distribution in units of $\rho_i$ (eq. [\ref{cordens}]) at the top and the magnetic potential
(eq. [\ref{streamfunc}]) at the bottom. Forty logarithmically spaced isocontour lines are shown
for the density; 40 linearly spaced contours for the magnetic potential.
The absolute nominal value for the steps between the isocontours of the logarithm of the
density is 0.098. The velocity and toroidal magnetic field are initially zero.
\label{f2}}
\figcaption[f3.ps]{Simulation A (without perturbation): (from top to bottom)
Density, poloidal velocity, poloidal magnetic field,
toroidal magnetic field and toroidal velocity for $\tau =400$. Forty logarithmically spaced
isocontour lines are shown for the density, and 40 linearly spaced contours for all the other
quantities.
The labels of the toroidal magnetic field and toroidal velocity denote 100
times their nominal values, in units of $B_i$ and $v_{Ki}$, respectively.
The absolute nominal values for the steps between the isocontours of the logarithm of the
density, toroidal magnetic field and toroidal velocity are  0.1, $0.034 B_i$ and
$0.0118 v_{Ki}$, respectively.
The maximum value of the poloidal velocity is equal to $0.64 v_{Ki}$.
We note the formation of a steady state jet. 
\label{f3}}
\figcaption[f4.ps]{Simulation B ($T=10$): (from top to bottom)
Density, poloidal velocity, poloidal magnetic field, toroidal magnetic field
and toroidal velocity for $\tau =400$. Forty logarithmically spaced
isocontour lines are shown for the density, and 40 linearly spaced contours for all the other
quantities. 
The labels of the toroidal magnetic field and toroidal velocity denote 100
times their nominal values.
The absolute nominal values for the steps between the isocontours of the logarithm of the
density, toroidal magnetic field and toroidal velocity are  0.08, $0.0368 B_i$ and
$0.0108 v_{Ki}$, respectively.
The maximum value of the poloidal velocity is equal to $0.64 v_{Ki}$. 
We note the formation of regularly spaced structures along the jet axis.
\label{f4}}
\figcaption[f5.ps]{Simulation C ($T=20$): (from top to bottom)
Density, poloidal velocity, poloidal magnetic field, toroidal magnetic field
and toroidal velocity for $\tau =400$, using a better resolution
($250,80$) uniform zones.
Forty logarithmically spaced
isocontour lines are shown for the density, and 40 linearly spaced contours for all the other
quantities. 
The labels of the toroidal magnetic field and toroidal velocity denote 100
times their nominal values.
The absolute nominal values for the steps between the isocontours of the logarithm of the
density, toroidal magnetic field and toroidal velocity are 0.094, $0.0346 B_i$ and
$0.0154 v_{Ki}$, respectively.
The maximum value of the poloidal velocity is equal to $0.61 v_{Ki}$. 
We note, again, the formation of regularly spaced structures along the jet
axis.
\label{f5}}
\figcaption[f6.ps]{Simulation D ($T=40$): (from top to bottom)
Density, poloidal velocity, poloidal magnetic field, toroidal magnetic field and
toroidal velocity for $\tau =400$. Forty logarithmically spaced
isocontour lines are shown for the density, and 40 linearly spaced contours for all the other
quantities. 
The labels of the toroidal magnetic field and toroidal velocity denote 100
times their nominal values.
The absolute nominal values for the steps between the isocontours of the logarithm of the
density, toroidal magnetic field and toroidal velocity are 0.094, $0.0342 B_i$
and $0.016 v_{Ki}$, respectively.
The maximum value of the poloidal velocity is equal to $0.67 v_{Ki}$. 
We note fragmentation of regularly spaced structures along the jet axis.
\label{f6}}
\figcaption[f7.ps]{Simulation E ($T=80$): (from top to bottom)
Density, poloidal velocity, poloidal magnetic field, toroidal magnetic field
and toroidal velocity for $\tau=400$. Forty logarithmically spaced
isocontour lines are shown for the density, and 40 linearly spaced contours for all the other
quantities. 
The labels of the toroidal magnetic field and toroidal velocity denote 100
times their nominal values.
The absolute nominal values for the steps between the isocontours of the logarithm of the
density, toroidal magnetic field and toroidal velocity are 0.096, $0.0344 B_i$
and $0.0152 v_{Ki}$, respectively.
The maximum value of the poloidal velocity is equal to $0.64 v_{Ki}$. 
We note more pronounced fragmentation of the regularly spaced structures along the jet axis. 
\label{f7}}
\figcaption[f8.ps]{Simulations A to E:
Density (solid line), axial velocity (dotted line), toroidal
velocity (dashed line) and toroidal magnetic field (dashed-dotted line) along
the $z$-axis at $r=5r_i$. The curves are normalized by their maximum values.
\label{f8}}
\figcaption[f9.ps]{Simulations A to E: Axial and radial mass fluxes (eq.[\ref{massflux}]).
Here we show the axial (solid line) and radial (dotted line) mass fluxes at $\tau =400$.
We calculate the axial mass flux, at each axial distance $z$, using the transverse area
$A=\pi r^2$ with $r=20r_i$. For the radial mass flux, we use the cylindrical surface area
$A=2 \pi r z$ with $r=20r_i$, at each axial distance $z$. The fluxes are given in units of
$\dot{m}_i=2 \pi r_i^2 \rho_i v_{Ki}$. The scale on the vertial axis represents a tenth of
the nominal value.
\label{f9}}
\figcaption[f10.ps]{Simulations A to E: The sonic Mach numbers
(dotted lines), Alfv\'enic Mach numbers (dashed-dotted lines) and magnetosonic Mach
numbers (solid lines) along the $z$-axis for all simulations at $\tau =400$ in
a cut at $r=5r_i$. For simulatations C to E, dashed-dotted lines denote $10^{-2}$ times the
nominal values of Alfv\'enic Mach numbers. 
 \label{f10}}
\figcaption[f11.ps]{Logarithmically spaced isocontour lines are shown for the density. On the
top we have the result $\beta_i = 0.1$. We obtained, in this case, smoothing of the
structures along the jet. The isocontours of the density slightly oscillate near the jet axis.
On the botton of the figure we used $\beta_i = 10$. Here we observe the formation of
structures displaced from the jet axis up to the distance of about $z = 35r_i$. The region
near the jet axis show irregular structures and is less well defined.
\label{f11}}
\end{document}